\documentstyle[11pt,newpasp,twoside,epsfig]{article}
\def\edcomment#1{\iffalse\marginpar{\raggedright\sl#1\/}\else\relax\fi}
\reversemarginpar

\newcommand{\ltsim}{\mbox{{\raisebox{-0.4ex}{$\stackrel{<}{{\scriptstyle\sim}}
$}}}}

\begin{document}
\title{A Search for Radio-loud Quasars within the Epoch of Reionization}
\author{Matt J.~Jarvis, Steve Rawlings, \& F.~Eugenio Barrio}
\affil{Astrophysics, Department of Physics, Keble Road, Oxford, OX1
3RH, UK}
\author{Gary J. Hill \& Amanda Bauer}
\affil{McDonald Observatory, University of Texas at Austin, 1
  University Station C1402, Austin, TX 78712-0259, USA}
\author{Steve Croft}
\affil{IGPP, Lawrence Livermore National Laboratory, L-413, 7000 East
  Ave., Livermore, CA 94550, USA}

\begin{abstract}
The Universe became fully reionized, and observable optically,
at a time corresponding to redshift $z \sim 6.5$,
so it is only by studying the HI and molecular absorption lines
against higher-redshift, radio-loud sources that one can hope to make detailed
studies of the earliest stages of galaxy formation. At present
no targets for such studies are known. In these proceedings we describe a survey which is underway to find radio-loud quasars at $z > 6.5$.

\end{abstract}

\section{Introduction}
After a long quest, the epoch of reionization has been
discovered as a protracted period reaching from $z \sim 20 \rightarrow
6.5$ (Kogut et
al. 2003; Becker et al. 2001).
Prior to $z \sim 6.5$, the Universe was optically opaque,
but galaxy formation was well underway.
Since it is impossible to study this `grey age' of the Universe optically,
observing the $z > 6.5$ Universe
is the primary science driver for the near-IR-optimized
NGST.
However, well before NGST flies, great progress can be expected
if radio and millimetre telescopes can be targeted on quasars
observed within the reionization epoch. Radio-loud
targets allow absorption studies that
can probe the evolving neutral and molecular content of the
high-$z$ Universe (e.g. Carilli, Gnedin, \& Owen 2002),
and radio HI absorption is the {\it only} way of probing
the neutral gas which goes
on to form stars. We could begin these studies with
current facilities (e.g.\ the GBT and GMRT), but we desperately need
measurements of space density
to influence the design of next-generation instruments like
LOFAR and the SKA. Unfortunately, there are currently no
known $z > 6.5$ radio-loud objects.

This is because such objects are rare --- a tiny
fraction ($\ll 1$ per cent) of the radio population.
Interest in pursuing them was
dampened by the claim of a much sharper cut-off in their
redshift distribution (Shaver et al.\ 1996) than earlier work
(Dunlop \& Peacock 1990) had suggested. Jarvis \& Rawlings (2000) and Jarvis et al.\ (2001)
have re-examined all the evidence concerning this
redshift cutoff,
obtaining results strongly favouring a fairly gradual decline with redshift.

\section{Design of the Survey}

Jarvis \& Rawlings (2000) emphasized the
care needed in sample selection and analysis.
The survey should be selected at low frequency to avoid losing the
highest-$z$ quasars because of the steepening of radio spectra at high
rest-frame frequencies.  The only low-frequency (325 MHz) survey with
the required depth and sky coverage is WENSS/WISH (Rengelink et
al.1997; De Breuck et al. 2002); its sky coverage and the Galactic
Plane forces us to the area shown in Fig.\ 1(a) (the upper declination
limit of $\delta = 60^{\circ}$ is due to consideration of the UKIRT
declination limit).

\begin{figure}[!h]
\plotone{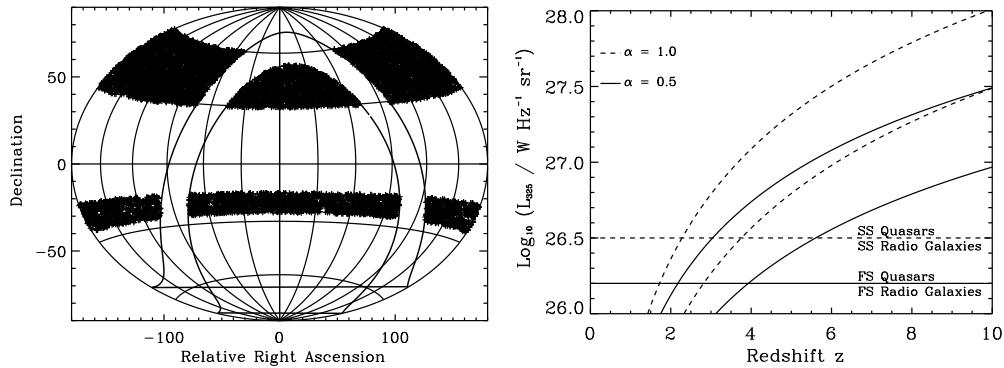}
\caption{({\it Left}) The areal coverage of our survey: there are 4559 flat-spectrum
sources in the north, and 1819 sources in the south. ({\it Right}) Tracks of $S_{325}=100$ mJy sources (upper lines) and
$S_{325}=30$ mJy sources
(lower lines) for steep-spectrum ($\alpha \sim 1$, SS)
sources (dashed) and flat-spectrum ($\alpha < 0.5$, FS)
sources (full). Note that, at a given
$S_{325}$, the change in luminosity of FS sources with redshift
is slower, meaning a larger ratio of $z=6.5$ to $z=5$ objects for
FS objects than SS objects (more so because the FS luminosity function
is flatter; Jarvis \& Rawlings 2000). The horizontal lines
show the luminosities at which the populations switch from
being dominated by quasars, to being dominated by radio galaxies
(Jarvis \& Rawlings 2000). The intersection of the full line
with the full curve at $z \sim 4$ shows that most $z>4$ FS
sources are quasars, while at $z < 4$, radio galaxies start to dominate
(McKean 2000). An important subtlety here is that because FS objects are
observed pole-on, a quasar nucleus should be seen when present, but a quasar
nucleus in powerful SS objects can easily be hidden by the obscuring torus.
}
\end{figure}

Analysis must account for distributions in the radio spectral shapes
of the population as highlighted by Jarvis et al. (2001). Our latest data-constrained model has been the basis of
simulations to test potential selection criteria for the FS
population, yielding optimum flux density and spectral index cuts.
Adopting these cuts there are $\approx 6500$ candidates in our
`quarter-sky' survey of which $\sim 0.1$ per cent will be at $z >
6.5$.
Our work allows us to quantify uncertainties in extrapolating
the number of $z > 6.5$ radio quasars from $z \sim 5$.
It is clear we should focus on FS rather than SS sources because: 

\begin{itemize}

\item {Their more favourable $k$-correction and luminosity function
[Fig.\ 1(b)] flattens their high-$z$ distribution.}

\item {At the highest $z$ they are mostly quasars
[Fig.\ 1(b)], so getting redshifts is relatively  easy
(c.f.\ impossibly-faint high-$z$ SS radio galaxies!).
}

\item {They are compact, so NVSS positions
yield reliable identifications.}

\item {Any relativistic beaming effects due to their
probable orientation along our line-of-sight means we probe much further down the radio luminosity function, towards the
bulk of the radio-loud quasar luminosity density (e.g. Willott et al. 2001).}

\end{itemize}

\section{Eliminating Low-redshift Interlopers}

There is a challenging, but tractable, sifting problem to eliminate both galaxies and quasars at $z < 6$. 
This is done in four steps.

(i) We have cross-correlated the radio sample with publicly available
all-sky optical and near-IR imaging, i.e.\ SDSS, POSS and 2MASS,
along with more general searches of known objects via the literature
and NED.
From this investigation we have optical IDs for
about 67\% of the objects in the northern sample (comprising quasars, low-redshift galaxies, BL Lac objects etc.). The remaining
$\approx 1500$ objects have no detectable optical emission, typically
to $R \sim 21.5$. 

(ii) We have
initiated deeper targeted observations in $R-$band of the remaining $\sim 1500$
northern sources with IGI at the 2.7~m at McDonald Observatory. 
Preliminary observations of $\sim 200$
sources down to $R \sim 23-23.5$ (depending on conditions) has shown
that of the remaining 1500 sources, approximately 60-70\% are
detected. These sources, generally extended objects which are
presumably $z \ltsim 2$ radio galaxies, are obviously not at $z >
6.5$, because like the $z \sim 6.5$ quasars already
known (e.g. Fan et al.~2003), these must have zero flux
below the redshifted Lyman limit.

(iii)  We use good-seeing near-IR imaging to find all the remaining
quasars, and eliminate all the remaining galaxies. [This works because, e.g.\ Fig.\ 1(b), the quasar fraction in the
FS population drops precipitously to lower redshifts
($z < 4$): indeed most FS sources at the depth of our survey 
will be galaxies (McKean et al.\ 2000); 10 nights on UKIRT have
already been allocated.].

(iv) We take near-IR spectra and find the $z > 6.5$ quasars. UKIRT-UIST has recently showed its capability in detecting
quasar broad-emission lines in the highest redshift quasar known to date
(Willott, McLure, \& Jarvis 2003) allowing an estimate of the
black-hole mass in this quasar via the broad MgII emission line. Lyman-$\alpha$ is more
than twice as bright as MgII in the composite SDSS quasar spectrum and the huge drop blueward of Lyman-$\alpha$ due to absorption by neutral hydrogen is also a very strong signature.
Therefore, identification would be relatively easy in $\sim 20$~min 
exposures on the Hobby-Eberly telescope with its planned J-band
extension to its low-resolution spectrograph and a huge amount of follow-up time will {\bf not} be required.


\section{Further Scientific Gains}

A survey of near-IR-selected, typically radio-quiet, $z > 6.5$ quasars
is to be pursued by the UKIDSS consortium using the UKIRT
(Warren 2002). They expect to find 
$\sim 10$ examples from $\sim 1 ~ \rm sr$. We expect to find
a similar number, despite the fact that only $\sim 10$\%
of quasars are radio-loud, because we cover 3-times the area, and
go significantly deeper in the near-IR.
Because radio selection is unbiased by dust,
spectroscopy/near-IR colours of our objects will help
assess the effects of dust obscuration on the UKIDSS selection function. At such early epochs, little
is currently known about the dust content of quasars and the consequent effect
on their near-IR colours.

\section{Summary}
We have outlined an efficient method to find radio-loud quasars within the epoch
of reionization. When discovered these objects will allow us to probe
intervening proto-galaxies and mini-haloes via absorption due to the hyperfine transition of
hydrogen at 21~cm. This will provide crucial information regarding the
build-up of the first galaxies as well as the quasar phenomena.

\acknowledgements

Work by SC is performed under the auspices of USDOE, NNSA by UC, LLNL under 
contract W-7405-Eng-48.

\references

Becker, R.L., et al. 2001, \aj, 122, 2850\\
Carilli, C.L., Gnedin, N.Y., \& Owen, F. 2002, \apj, 577, 22\\
De Breuck, C., Tang, Y. de Bruyn, A.G., R\"ottgering, H.J.A., \& van Breugel, W. 2002, \aap, 394, 59\\
Dunlop, J.S., \& Peacock, J.A. 1990, \mnras, 247, 19 \\
Fan, X., et al. 2003, \aj, 125, 1649\\
Jarvis, M.J., \& Rawlings S. 2000, \mnras, 319, 121\\
Jarvis, M.J., et al. 2001, \mnras, 327, 907\\
Kogut et al. 2003, ApJ, submitted (astro-ph/0302213)\\
McKean et al. 2000, http://www.jb.man.ac.uk/$\sim$jmckean/publications.html.\\
Rengelink, R.B., Tang, Y., de Bruyn, A.G., Miley, G.K., Bremer, M.N., R\"ottgering, H.J.A., \& Bremer, M.A.R. 1997, \aaps, 124, 259\\
Shaver, P.A., Wall, J.V., Kellermann, K.I., Jackson, C.A., \& Hawkins, M.R.S. 1996, Nature, 384, 439\\
Warren, S. 2002, http://www.ukidss.org/sciencecase/sciencecase.html.\\
Willott, C.J., McLure, R.J., \& Jarvis, M.J. 2003, \apj, 587, 15\\
Willott, C.J., Rawlings, S., Blundell, K.M., Lacy, M., \& Eales, S.A. 2001, \mnras, 322, 536\\

\end{document}